\documentclass[prx,letterpaper,nobalancelastpage,twocolumn,superscriptaddress,nofootinbib]{revtex4-1}

\usepackage{graphicx}
\usepackage{amsmath}
\usepackage{bbold}
\usepackage{amssymb}
\usepackage[english]{babel}
\usepackage{color}
\usepackage[version=4]{mhchem}
\usepackage[hidelinks]{hyperref}
\usepackage{dsfont}

\setcitestyle{super}

\usepackage{soul}
\newcommand{\mc}[1]{\mathcal{#1}}

\newcommand{\ket}[1]{| #1 \rangle}
\newcommand{\bra}[1]{\langle #1 |}

\newcommand{\Caltech}{Division of Physics, Mathematics and Astronomy, California Institute of Technology, Pasadena, CA 91125, USA}
\newcommand{\JPL}{Jet Propulsion Laboratory, California Institute of Technology, Pasadena, CA 91109, USA}

\newcommand{\tr}[1]{\mathrm{tr}\left\{#1\right\}}
\newcommand{\lr}[1]{\left( #1 \right)}

\usepackage{caption}

\DeclareCaptionLabelSeparator{bar}{ \textbf{\textbar}~}
\begin{document}

\title{High-fidelity entanglement and detection of alkaline-earth Rydberg atoms}
\author{Ivaylo S. Madjarov}\thanks{These authors contributed equally to this work}
\affiliation{\Caltech}
\author{Jacob P. Covey}\thanks{These authors contributed equally to this work}
\affiliation{\Caltech}
\author{Adam L. Shaw}
\affiliation{\Caltech}
\author{Joonhee Choi}
\affiliation{\Caltech}
\author{Anant Kale}
\affiliation{\Caltech}
\author{Alexandre Cooper}\altaffiliation{Permanent address: Institute for Quantum Computing, University of Waterloo, 200 University Ave West, Waterloo, Ontario, Canada}
\author{Hannes Pichler}
\affiliation{\Caltech}
\author{Vladimir Schkolnik}
\affiliation{\JPL}
\author{Jason R. Williams}
\affiliation{\JPL}
\author{Manuel Endres}\email{mendres@caltech.edu}
\affiliation{\Caltech}

\begin{abstract}
Trapped neutral atoms have become a prominent platform for quantum science, where entanglement fidelity records have been set using highly-excited Rydberg states. However, controlled two-qubit entanglement generation has so far been limited to alkali species, leaving the exploitation of more complex electronic structures as an open frontier that could lead to improved fidelities and fundamentally different applications such as quantum-enhanced optical clocks. Here we demonstrate a novel approach utilizing the two-valence electron structure of individual alkaline-earth Rydberg atoms. We find fidelities for Rydberg state detection, single-atom Rabi operations, and two-atom entanglement surpassing previously published values. Our results pave the way for novel applications, including programmable quantum metrology and hybrid atom-ion systems, and set the stage for alkaline-earth based quantum computing architectures. 
\end{abstract}

\maketitle
Recent years have seen remarkable advances in generating strong, coherent interactions in arrays of neutral atoms through excitation to Rydberg states, characterized by large electronic orbits~\cite{Saffman2010,Browaeys2016,Saffman2016,Browaeys2020}. This has led to profound results in quantum science applications, such as quantum simulation~\cite{Schauss2015,Labuhn2016,Bernien2017,Browaeys2020} and quantum computing~\cite{Browaeys2016,Saffman2016,Jau2016,Levine2018,Graham2019,Levine2019}, including a record for two-atom entanglement for neutral atoms~\cite{Levine2018}. Furthermore, up to 20-qubit entangled states have been generated in Rydberg arrays~\cite{Omran2019}, competitive with results in trapped ions~\cite{Monz2011} and superconducting circuits~\cite{Song2019}.  Many of these developments were fueled by novel techniques for generating reconfigurable atomic arrays~\cite{Barredo2016,Endres2016,Kumar2018} and mitigation of noise sources~\cite{Levine2018,deLeseleuc2018}. While previous Rydberg-atom-array experiments have utilized alkali species, atoms with a more complex level structure, such as alkaline-earth atoms (AEAs)~\cite{DeSalvo2016,Gaul2016,Lochead2013,Norcia2018b,Cooper2018a,Saskin2018} commonly used in optical lattice clocks~\cite{Ludlow2015}, provide new opportunities for increasing fidelities and accessing fundamentally different applications, including Rydberg-based quantum metrology~\cite{Gil2014,Kessler2014,Kaubruegger2019}, quantum clock networks~\cite{Komar2014}, and quantum computing schemes with optical and nuclear qubits~\cite{Daley2008,Gorshkov2009}.

\begin{figure}[t!]
	\centering
	\includegraphics[width=8.5cm]{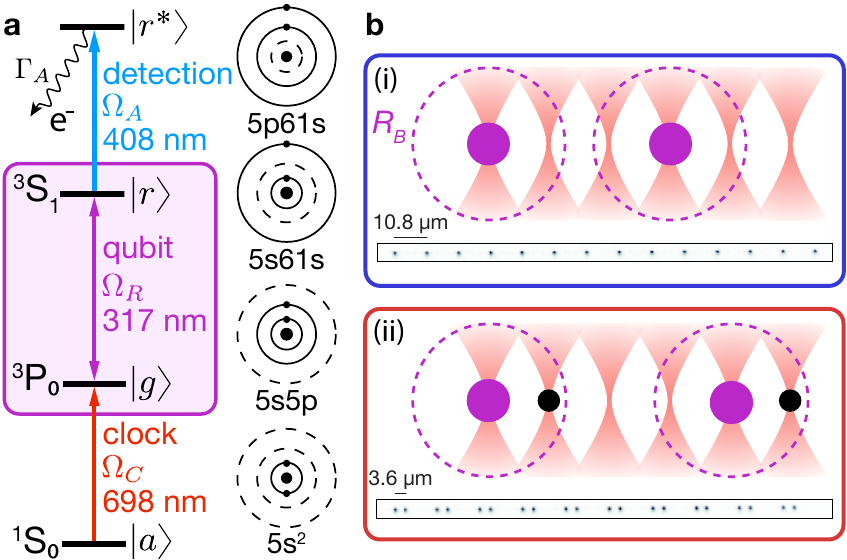}
	\caption{\textbf{Population and detection of Rydberg states in non-interacting and interacting configurations.} $\textbf{a}$, The relevant level structure (left), and electronic configuration (right) for strontium$-88$. The Rydberg-ground state qubit is defined by a metastable `clock' state $|g\rangle$ and the 5s61s $^3$S$_1$ $m_\text{J}=0$ Rydberg state $|r\rangle$ (highlighted with a purple box), which we detect by driving to an auto-ionizing 5p61s state $|r^*\rangle$. The clock state $|g\rangle$ is initialized from the absolute ground state $|a\rangle$. \textbf{b}, We use atom-by-atom assembly in optical tweezers to prepare an effectively non-interacting configuration ((i), blue box and data-points throughout) and a strongly Rydberg-blockaded pair configuration ((ii), red box and data-points throughout)~\cite{Bernien2017}. The blockade radius $R_B$, where two-atom excitation is suppressed, is indicated by a dashed circle. Throughout, purple and black circles indicate $|r\rangle$ and $|g\rangle$ atoms, respectively. The Rydberg, auto-ionization, and clock beams all propagate along the axis of the atom array and address all atoms simultaneously. Averaged fluorescence images of atoms in configurations (i) and (ii) are shown. See Methods and Supplementary Information for further detail. }\vspace{-0.5cm}
	\label{FigSchematicMain}
\end{figure} 

Here we demonstrate such a novel Rydberg array architecture based on AEAs, where we utilize the two-valence electron structure for single-photon Rydberg excitation from a meta-stable clock state as well as auto-ionization detection of Rydberg atoms  (Fig.~\ref{FigSchematicMain}). We find leading fidelities for Rydberg state detection, ground- to Rydberg-state coherent operations, and Rydberg-based two-atom entanglement (Table~\ref{table:main}). More generally, our results constitute the highest reported two-atom entanglement fidelities for neutral atoms~\cite{Kaufman2015,Welte2018,Levine2018} as well as a proof-of-principle for controlled two-atom entanglement between AEAs. We further demonstrate a high-fidelity entanglement operation with optical traps kept on, an important step for gate-based quantum computing~\cite{Saffman2010,Browaeys2016,Saffman2016,Jau2016,Levine2018,Graham2019,Levine2019}. As detailed in the outlook section, our results open up a host of new opportunities for quantum metrology and computing as well as for optical trapping of ions.

\begin{table}[t!]
\caption{\textbf{Uncorrected and SPAM-corrected fidelities for single-atom and Rydberg-blockaded pulses.} The `T' indicates settings where the tweezers are on during Rydberg excitation.}
\centering
\renewcommand{\arraystretch}{1.5} 
\begin{tabular}{p{4cm}{c}{c}{c}{c}}
\hline\hline
Quantity&Uncorrected&SPAM-corrected\\
\hline
Single-atom $\pi$-pulse & $0.9951(9)$ & $0.9967(9)$ \\
Single-atom $2\pi$-pulse & $0.9951(9)$ & $0.998(1)$ \\
Blockaded $\pi$-pulse & $0.992(2)$ & $0.996(2)$ \\
Blockaded $2\pi$-pulse & $0.992(2)$ & $0.999(2)$ \\
Blockaded $\pi$-pulse, T & $0.992(2)$ & $0.996(2)$ \\
Blockaded $2\pi$-pulse, T & $0.987(2)$ & $0.994(3)$ \\
Bell state fidelity & $\geq0.980(3)$ & $\geq0.991(4)$ \\
Bell state fidelity, T & $\geq0.975(3)$ & $\geq0.987(4)$ \\
\hline\hline
\end{tabular}\vspace{-0.5cm}
\label{table:main}
\end{table}

Our experimental system~\cite{Cooper2018a,Covey2019a,Madjarov2019} combines various novel key elements: First, we implement atom-by-atom assembly in reconfigurable tweezer arrays~\cite{Barredo2016,Endres2016} for AEAs (Fig.~\ref{FigSchematicMain}b). Second, we sidestep the typical protocol for two-photon excitation to S-series Rydberg states, which requires significantly higher laser power to suppress intermediate state scattering, by transferring atoms to the long-lived $^3$P$_0$ clock state $\ket{g}$~\cite{Ludlow2015,Covey2019a,Madjarov2019,Norcia2019}. We treat $\ket{g}$ as an effective ground state from which we apply single-photon excitation to a $^3$S$_1$ Rydberg state $\ket{r}$~\cite{Gil2014}. Third, instead of relying on loss through tweezer anti-trapping as in alkali systems, we employ a rapid auto-ionization scheme for Rydberg state detection. In contrast to earlier implementations of auto-ionization detection in bulk gases~\cite{Lochead2013}, we image remaining neutral atoms~\cite{Covey2019a} instead of detecting charged particles. 

More generally, our findings improve the outlook for Rydberg-based quantum computing~\cite{Saffman2010,Browaeys2016,Saffman2016,Jau2016,Levine2018,Graham2019,Levine2019}, optimization~\cite{Pichler2018}, and simulation~\cite{Schauss2015,Labuhn2016,Bernien2017,Browaeys2020}. These applications all rely on high fidelities for preparation, detection, single-atom operations, and entanglement generation for which we briefly summarize our results: we obtain a state preparation fidelity of $0.997(1)$ through a combination of coherent and incoherent transfer~\cite{SMPRL2020}. The new auto-ionization scheme markedly improves the Rydberg state detection fidelity to $0.9963-0.9996$~\cite{Levine2018,Omran2019,SMPRL2020}. We also push the limits of single and two-qubit operations in ground- to Rydberg-state transitions~\cite{Levine2018,Labuhn2016,Graham2019,Omran2019}. For example, we find $\pi$-pulse fidelities of $0.9951(9)$ without correcting for state preparation and measurement (SPAM) and $0.9967(9)$ if SPAM correction is applied~\cite{SMPRL2020}. Finally, using a conservative lower-bound procedure, we observe a two-qubit entangled Bell state fidelity of $\geq0.980(3)$ and $\geq0.991(4)$ without and with SPAM correction, respectively. We note that all values are obtained on average and for parallel operation in arrays of 14 atoms or 10 pairs for the non-interacting or pair-interacting case, respectively.

\begin{figure}[t!]
	\centering
	\includegraphics[width=6.0cm]{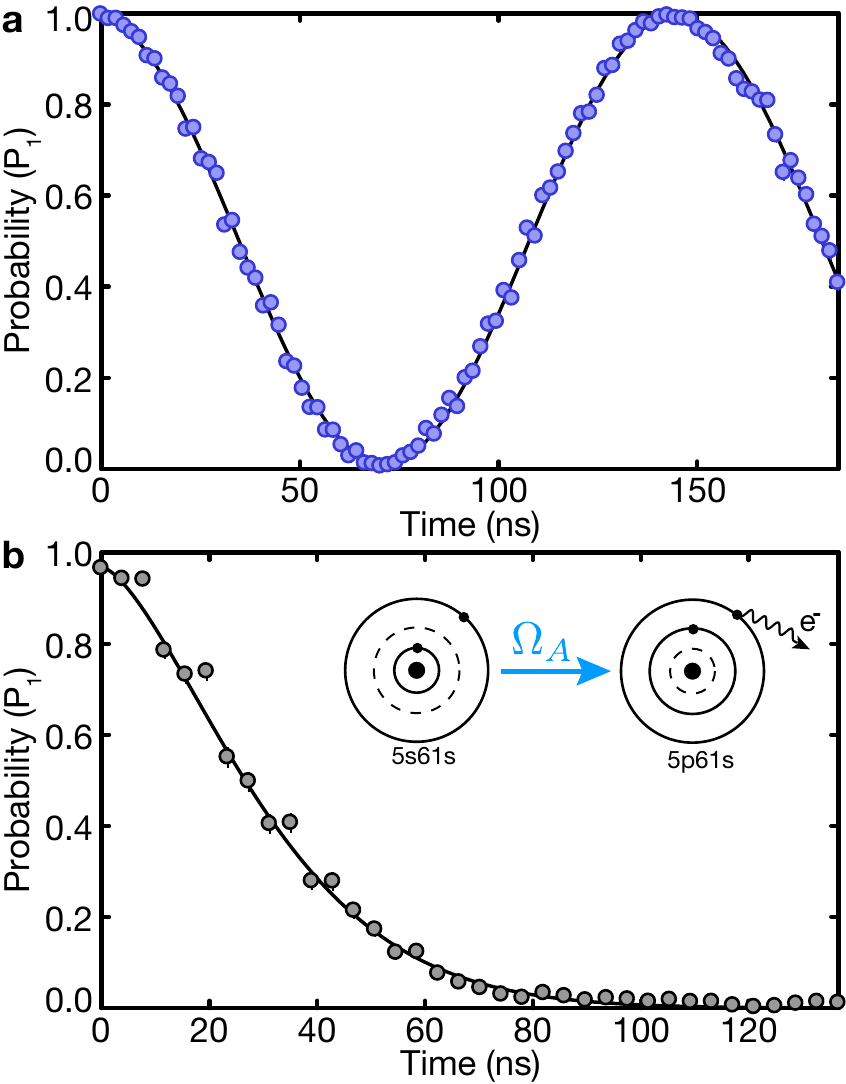}
	\caption{\textbf{Rabi oscillations and auto-ionization.} \textbf{a}, Array-averaged probability $P_1$ of detecting an atom after a resonant Rydberg pulse and subsequent auto-ionization as a function of Rydberg pulse time, showing high-contrast Rabi oscillations with frequency $\Omega_R=2\pi\times6.80(2)$ MHz. The auto-ionization pulse time is fixed to 5 $\mu$s. \textbf{b}, $P_1$ as a function of auto-ionization pulse time at a fixed Rydberg pulse time of $70$ ns corresponding to a $\pi$-pulse (followed by a second $\pi$-pulse). The solid line is a fit to a Gaussian, phenomenologically chosen to capture the finite switch-on time of the auto-ionization beam~\cite{SMPRL2020}. Inset: illustration of the auto-ionization process. In both \textbf{a} and \textbf{b}, data is uncorrected and averaged over $\approx40-100$ experimental cycles per timestep and over an array of approximately 14 atoms. Error bars indicate a $1\sigma$ binomial confidence interval.}\vspace{-0.5cm}
	\label{FigShortRabiAI}
\end{figure} 

\begin{figure*}[t!]
	\centering
	\includegraphics[width=0.8\textwidth]{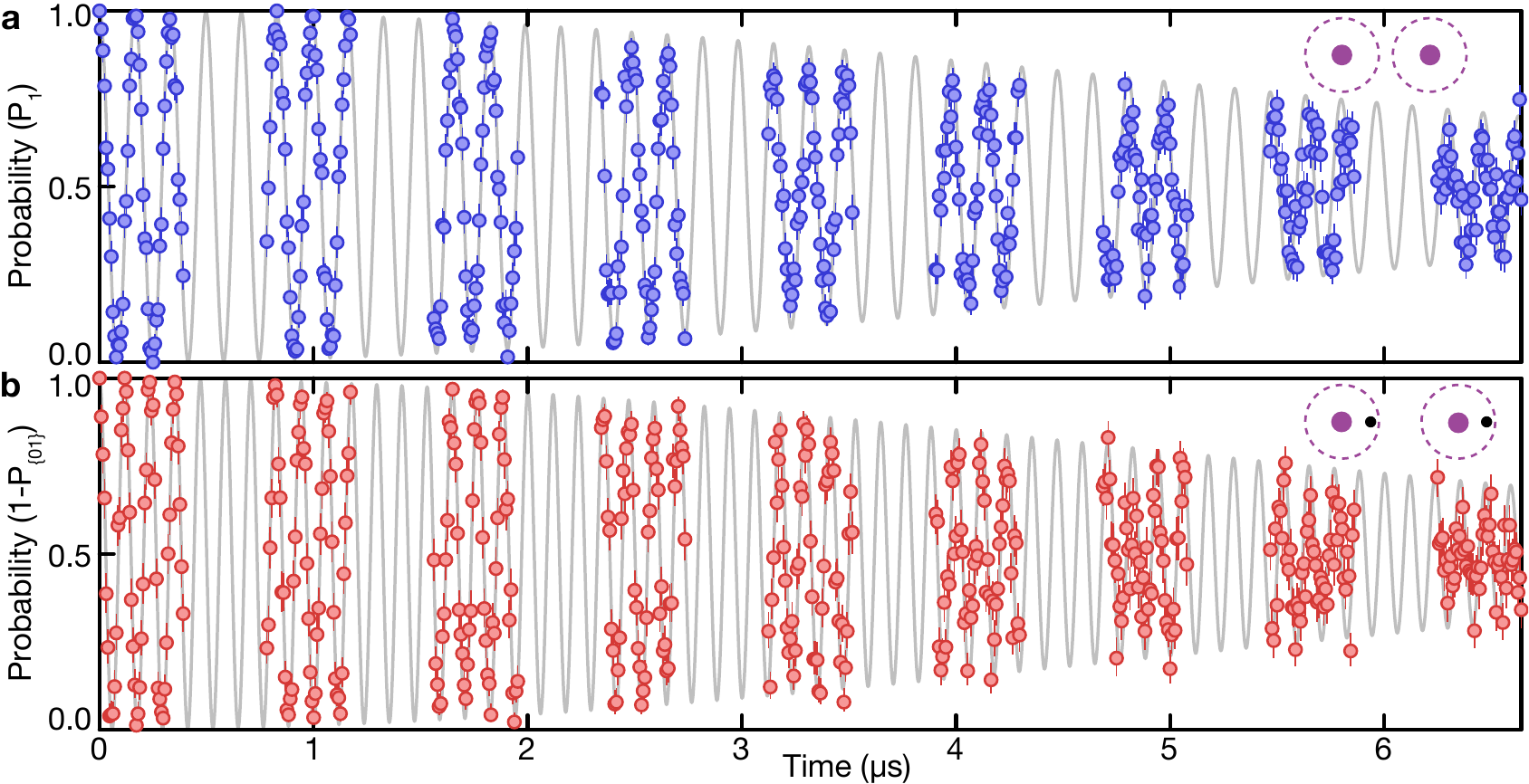}
	\caption[width=16.0cm]{\textbf{Long-time Rabi oscillations for single and blockaded atoms.} \textbf{a}, Array-averaged Rabi oscillations for the non-interacting configuration (i), depicted by the inset. We operate with $\Omega_R=2\pi\times6.0$ MHz. By fitting with a Gaussian profile, we find a $1/e$ coherence of $\approx42$ cycles. \textbf{b}, Same as in \textbf{a} but for the blockaded configuration (ii), depicted by the inset. We plot $1-P_{\{01\}}$, where $P_{\{01\}}$ is the array-averaged symmetrized probability of detecting one atom of an initial pair (and not both). We observe a blockade-enhanced Rabi frequency of $\tilde{\Omega}_R=2\pi\times8.5$ MHz. We find a $1/e$ coherence of $\approx 60$ cycles. In both \textbf{a} and \textbf{b}, data is uncorrected and averaged over $\approx10$ experimental cycles per timestep and over an array of approximately 14 atoms in \textbf{a} or 10 pairs in \textbf{b}. Error bars indicate a $1\sigma$ binomial confidence interval.}
	\label{FigLongRabi}\vspace{-0.5cm}
\end{figure*}

We begin by analyzing short-time Rabi oscillations between $\ket{g}$ and $\ket{r}$ (Fig.~\ref{FigShortRabiAI}a) and the auto-ionization detection scheme (Fig.~\ref{FigShortRabiAI}b) in an essentially non-interacting atomic configuration ((i) in Fig.~\ref{FigSchematicMain}b). To detect atoms in $|r\rangle$ we excite the core valence-electron from a 5s to a 5p level, which then rapidly auto-ionizes the Rydberg electron (inset of Fig.~\ref{FigShortRabiAI}b)~\cite{SMPRL2020}. The ionized atoms are dark to subsequent detection of atoms in $\ket{g}$~\cite{Covey2019a}, providing the means to distinguish ground and Rydberg atoms. 

We use a $|g\rangle\leftrightarrow|r\rangle$ Rabi frequency of $\Omega_R\approx2\pi\times6-7$ MHz throughout, and observe Rabi oscillations with high contrast at a fixed auto-ionization pulse length (Fig.~\ref{FigShortRabiAI}a, Table I). To quantify the auto-ionization detection, we perform a $\pi$-pulse on $|g\rangle\leftrightarrow|r\rangle$, then apply an auto-ionization pulse for a variable duration (Fig.~\ref{FigShortRabiAI}b), and then perform a second $\pi$-pulse on $|g\rangle\leftrightarrow|r\rangle$ before measurement. The detected population decreases to zero with a $1/e$ time of $\tau_A=35(1)$ ns. We can compare $\tau_A$ to the lifetime of $|r\rangle$, which is estimated to be $\tau_{|r\rangle}\approx80$ $\mu$s~\cite{Vaillant2012}, placing an upper bound on the $\ket{r}$-state detection efficiency of  $0.9996(1)$. A lower bound comes from the measured $\pi$-pulse fidelity of $0.9963(9)$ corrected for preparation and ground state detection errors. These limits can be increased with higher laser power and faster switching~\cite{SMPRL2020}.

To probe our longer-time coherence, we drive the Rydberg transition for as long as 7 $\mu$s (Fig.~\ref{FigLongRabi}a). The decay of the contrast on longer timescales is well modeled by a Gaussian profile of the form $\mathcal{C}(t)=\mathcal{C}_0\text{exp}(-t^2/\tau_C^2)$. We find that $\tau_C\approx7$ $\mu$s is consistent with our data, and corresponds to a $1/e$ coherence of $\approx42$ cycles. To our knowledge, this is the largest number of coherent ground-to-Rydberg cycles that has been published to date~\cite{Levine2018,Levine2019}. Limitations to short- and long-term coherence are discussed and modeled in detail in Ref.~\cite{SMPRL2020}. The main contributing factors are laser intensity and phase noise (which both can be improved upon with technical upgrades, such as cavity filtering of phase noise~\cite{Levine2018}), and finite Rydberg state lifetime.

\begin{figure}[t!]
	\centering
	\includegraphics[width=6cm]{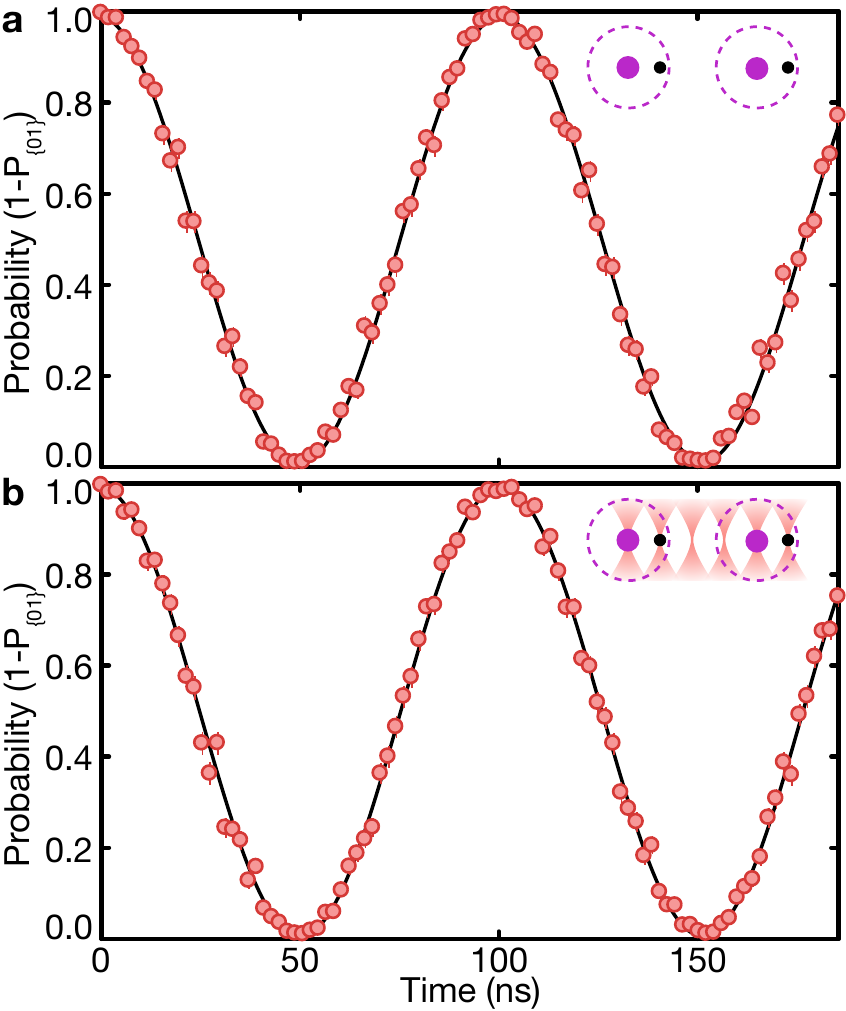}
	\caption{\textbf{Short-time Rydberg-blockaded Rabi oscillations with tweezers \textit{off} and \textit{on}.} \textbf{a}, Short-time Rabi-oscillations for the blockade configuration (ii) with the traps \textit{off}, depicted by the inset. \textbf{b}, Same as in \textbf{a} but with tweezers \textit{on} during Rydberg interrogation with a $|g\rangle$-state depth of $\text{U}/h\approx0.94$ MHz. The blockade-enhanced Rabi frequency is $\tilde{\Omega}_R=2\pi\times9.8$ MHz.  In both \textbf{a} and \textbf{b}, data is uncorrected and averaged over $\approx50-100$ experimental cycles per timestep and over an array of approximately 10 pairs. Error bars indicate a $1\sigma$ binomial confidence interval.}
	\label{FigRydbergBlockade}
\end{figure}

We now turn to the pair-interacting configuration ((ii) in Fig.~\ref{FigSchematicMain}b) to study blockaded Rabi oscillations~\cite{Saffman2010,Levine2018}. For an array spacing of $3.6$ $\mu$m, we anticipate an interaction shift of $V_B\approx2\pi\times130$ MHz for the $n=61$ Rydberg state in the $^3$S$_1$ series~\cite{Vaillant2012}. In this configuration, simultaneous Rydberg excitation of closely-spaced neighbors is strongly suppressed, and an oscillation between $|gg\rangle$ and the entangled $W$-Bell-state $|W\rangle=(|gr\rangle+e^{i\phi}|rg\rangle)/\sqrt{2}$ is predicted with a Rabi frequency enhanced by a factor of $\sqrt{2}$~\cite{Saffman2010}, as observed in our data. We show our results for long-term coherent oscillations in Fig.~\ref{FigLongRabi}b and find a $1/e$ coherence time corresponding to $\approx60$ cycles. Results for short-term oscillations are shown in Fig.~\ref{FigRydbergBlockade}a and the fidelity values are summarized in Table~\ref{table:main}.

We now estimate the Bell state fidelity associated with a two-atom (blockaded) $\pi$-pulse. While parity oscillations provide a standard metric for entanglement fidelity~\cite{Levine2018}, they require site-resolved laser addressing. We leave this technique for future work, and instead provide a lower bound for the Bell state fidelity based on measured populations at the (blockaded) $\pi$-time and a lower bound on the purity of the two-atom state. The latter is obtained by measuring the atomic populations at the (blockaded) $2\pi$ time, under the assumption that the purity does not increase between the $\pi$ and the $2\pi$ time. For a detailed discussion and analysis of this bound and the validity of the underlying assumptions, see Ref.~\cite{SMPRL2020}. With this approach, we find uncorrected and SPAM-corrected lower bounds on the Bell state fidelity of $0.980(3)$ and $0.991(4)$, respectively (Table~\ref{table:main}). 

We note that all preceding results were obtained with the tweezers switched \textit{off} during Rydberg excitation. The potential application of Rydberg gates to large circuit depth quantum computers motivates the study of blockade oscillations with the tweezers \textit{on}. In particular, we foresee challenges for sequential gate-based platforms if tweezers must be turned off during each operation to achieve high fidelity. In systems implementing gates between the absolute ground and clock states for example, blinking traps on and off will eventually lead to heating and loss, ultimately limiting the number of possible operations. Furthermore, while individual tweezer blinking is possible in one dimension, the prospects for blinking individual tweezers in a two dimensional array are unclear: a two-dimensional array generated by crossed acousto-optic deflectors cannot be blinked on the level of a single tweezer, and one generated by a spatial light modulator cannot be blinked fast enough to avoid loss. Repulsive traps such as interferometrically-generated bottles~\cite{Barredo2019} or repulsive lattices~\cite{Graham2019} have been developed in lieu of standard optical tweezer arrays~\cite{Endres2016,Barredo2016} in part to help maintain high-fidelity operations while keeping traps on. 
 
Despite finding that our Rydberg state is anti-trapped (with a magnitude roughly equal to that of the ground state trapping) at our clock-magic wavelength of $\lambda_T = 813.4$~nm~\cite{SMPRL2020}, we observe high-fidelity entanglement even when the tweezers remain on during Rydberg interrogation. Certain factors make this situation favorable for alkaline-earth atoms. One is the ability to reach lower temperatures using narrow-line cooling, which suppresses thermal dephasing due to trap light shifts. Furthermore, a lower temperature allows for ramping down of tweezers to shallower depths before atoms are lost, further alleviating dephasing. Finally, access to higher Rabi frequencies provides faster and less light-shift-sensitive entangling operations.

We study short-time blockaded Rabi oscillations both with the tweezers switched \textit{off} (Fig.~\ref{FigRydbergBlockade}a) and left \textit{on} (Fig.~\ref{FigRydbergBlockade}b). We find similar fidelities for the $\pi$- and $2\pi$-pulses in both cases (Table~\ref{table:main}). Further, we estimate a lower bound for the Bell state fidelity in the tweezer \textit{on} case, and find uncorrected and corrected values of $\geq0.975(3)$ and $\geq0.987(4)$, respectively. We expect further improvements in shorter-wavelength tweezers for which the Rydberg states of AEAs are trapped~\cite{Mukherjee2011}, and our observations show promise for Rydberg-based quantum computing in a standard tweezer array~\cite{Endres2016,Barredo2016}.

Our work bridges the gap between the fields of Rydberg atom arrays and optical clocks~\cite{Ludlow2015}, opening the door to Rydberg-based quantum-enhanced metrology~\cite{Gil2014,Kessler2014}, including the programmable generation of spin-squeezed states~\cite{Kaubruegger2019} in recently demonstrated tweezer clocks~\cite{Madjarov2019,Norcia2019}, and quantum clock networks~\cite{Komar2014}. Further, the demonstrated entangling operations provide a mechanism for two-qubit gates in AEA-based quantum computation and simulation architectures leveraging optical and nuclear qubits~\cite{Daley2008,Gorshkov2009}. More generally, the observed entanglement fidelities could enable gate fidelities for long-lived ground states approaching fault-tolerant error correction thresholds~\cite{Knill2005}. In addition, the high Rydberg- and ground-state detection-fidelities could play an important role in applications based on sampling from bit-string probability distributions~\cite{Pichler2018,Arute2019}. Finally, by auto-ionizing the Rydberg electron with high fidelity and noting that we expect the remaining ion to stay trapped, we have found a potential new approach to the optical trapping of ions~\cite{Karpa2013,Huber2014} in up to three dimensional arrays~\cite{Barredo2018,Kumar2018}. Such a platform has been proposed for ion-based quantum computing~\cite{Cirac2000} as well as for hybrid atom-ion systems~\cite{Engel2018,Mukherjee2019,Langin2019}. \\

\noindent
\textit{Note added.---}Recently, we became aware of work in ytterbium tweezer arrays demonstrating trapping of Rydberg states~\cite{Wilson2019}. \\ 
\\

\section*{Methods}

We briefly summarize the relevant features of our $^{88}$Sr experiment~\cite{Cooper2018a,Covey2019a,Madjarov2019}. We employ a one-dimensional array of 43 tweezers spaced by 3.6 $\mu$m. Atoms are cooled close to the tranverse motional ground state using narrow line cooling~\cite{Covey2019a,Norcia2019,Madjarov2019}, with an average occupation number of $\bar{n}_r\approx0.3$ ($T_r\approx2.5$ $\mu$K), in tweezers of ground-state depth $\text{U}_0\approx k_B\times450$ $\mu$K~$\approx h\times9.4$ MHz with a radial trapping frequency of $\omega_r\approx 2\pi\times78$ kHz.

For state preparation (Fig. 1a), we drive from the 5s$^2$ $^1$S$_0$ absolute ground state (labeled $|a\rangle$) to the 5s5p $^3$P$_0$ clock state (labeled $|g\rangle$) with a narrow-line laser~\cite{Madjarov2019}, reaching Rabi frequencies of $\Omega_C\approx2\pi\times3.5$ kHz in a magnetic field of $\approx710$ G ~\cite{Taichenachev2006a,Barber2006} (otherwise set to $\approx71$~G for the entire experiment). We populate $|g\rangle$ with a $\pi$-pulse reaching a loss-corrected fidelity of $0.986(2)$, which we supplement with incoherent pumping (after adiabatically ramping down the tweezer depth to $\text{U}_F=\text{U}_0/10$) to obtain a clock state population without and with loss correction of $0.997(1)$ and $0.998(1)$, respectively. This value is similar to, or higher than, the state preparation fidelities achieved with alkali atoms~\cite{Wang2016,Omran2019,Levine2019,Graham2019}. 

We treat the long-lived state $|g\rangle$ as a new ground state, from which we drive to the 5s61s $^3$S$_1$, $m_\text{J}=0$ Rydberg state (labeled $|r\rangle$). The $|g\rangle\leftrightarrow|r\rangle$ Rydberg transition occurs at a wavelength of $\lambda_\text{R}=316.6$ nm and we use a $1/e^2$ beam radius of $18(1)$ $\mu$m. We readily achieve a $|g\rangle\leftrightarrow|r\rangle$ Rabi frequency of $\Omega_R\approx2\pi\times6-7$ MHz, corresponding to $\approx30$ mW, and up to $\Omega_R\approx2\pi\times13$ MHz with full optimization of the laser system and beam path. To detect atoms in $|r\rangle$ we drive the strong transition to 5p$_{3/2}$61s$_{1/2}$ ($J=1, m_\text{J}=\pm1$), labelled $|r^*\rangle$. This transition excites the core ion, which then rapidly auto-ionizes the Rydberg electron. The ionized atoms are dark to subsequent detection of atoms in $\ket{g}$ with the high-fidelity scheme described in Ref.~\cite{Covey2019a}, providing the means to distinguish ground and Rydberg atoms. We switch off the ramped-down tweezers during the Rydberg pulse~\cite{Labuhn2016,Bernien2017}, after which we apply an auto-ionization pulse while rapidly increasing the depth back to $\text{U}_0$ for subsequent read-out.

The Rydberg and clock laser beams are linearly polarized along the magnetic field axis, and the auto-ionization beam is linearly polarized perpendicular to the magnetic field axis. Accordingly, we excite to auto-ionizing states with $m_\text{J}=\pm1$. The tweezers are linearly polarized along the axis of propagation of the Rydberg, clock, and auto-ionization beams -- perpendicular to the magnetic field axis.

\section*{Acknowledgements}

We acknowledge discussions with Chris Greene and Harry Levine as well as funding provided by the Institute for Quantum Information and Matter, an NSF Physics Frontiers Center (NSF Grant PHY-1733907), the NSF CAREER award (1753386), the AFOSR YIP (FA9550-19-1-0044), the Sloan Foundation, and Fred Blum. Research was carried out at the Jet Propulsion Laboratory and the California Institute of Technology under a contract with the National Aeronautics and Space Administration and funded through the President’s and Director’s Research and Development Fund (PDRDF). JPC acknowledges support from the PMA Prize postdoctoral fellowship, and JC acknowledges support from the IQIM postdoctoral fellowship. HP acknowledges support by the Gordon and Betty Moore Foundation. AK acknowledges funding from the Larson SURF fellowship, Caltech Student-Faculty Programs. \\

\section*{Author Contributions}

M.E. conceived the idea and initiated the study. I.M., J.P.C., A.S., J.C., A.C., and V.S. designed and carried out the experiments. I.M., J.P.C., A.S., J.C., A.K., and H.P. performed theory and simulation work. I.M., J.P.C., A.S., J.C. contributed to data analysis. I.M., J.P.C., A.S., J.C., and M.E. contributed to writing the manuscript and supplementary information. J.P.C., J.W., and M.E. supervised and guided this work. I.M. and J.P.C. contributed equally to this work.\\

\section*{Data availability}

The data that support the findings of this study are available from the corresponding author upon reasonable request.

\section*{Competing interests}

The authors declare no competing interests.  
\clearpage
\begin{center}
\textbf{\Large Supplementary Information}
\end{center}

\setcounter{section}{0}

\twocolumngrid

\renewcommand\appendixname{Appendix}
\renewcommand\thesection{\Alph{section}}
\renewcommand\thesubsection{\arabic{subsection}}

\vspace{-3mm}
\section{State preparation}\label{SecStatePrep}
The ground state $|g\rangle$ of our Rydberg qubit is the 5s5p $^3$P$_0$ metastable clock state of $^{88}$Sr. We populate this state in two stages: first, most atoms are transferred via a coherent $\pi$-pulse on the clock transition. Thereafter, any remaining population is transferred via incoherent pumping. 

In our regime where the Rabi frequency of the clock transition ($\Omega_C \approx 2\pi \times 3.5$~kHz) is significantly smaller than the trapping frequency ($\omega_r \approx 2\pi \times 78$~kHz), coherent driving is preferable to incoherent pumping because it preserves the motional state of an atom, i.e., it does not cause heating.  However, atomic temperature, trap frequency, trap depth, and beam alignment contribute to the transfer infidelity of coherent driving.  Although we drive the clock transition on the motional carrier in the sideband resolved regime, thermal dephasing still plays an important role. Particularly, each motional state has a distinct Rabi frequency, a thermal ensemble of which leads to dephasing~\cite{Madjarov2019}. This thermal dephasing is less severe at higher trapping frequencies; however, this can only be achieved in our system by using deeper traps, which would also eventually limit transfer fidelity because of higher rates of Raman scattering out of the clock state. We therefore perform coherent transfer initially in deeper traps ($\approx 450~\mu\text{K}$), followed immediately by an adiabatic rampdown to one-tenth of that depth. Finally, precise alignment of the clock beam to the tight, transverse axis of the tweezer is important to ensure that no coupling exists to axial motion, which has a much lower trap frequency and thus suffers more thermal dephasing than the transverse direction. 

The remaining population is transferred by simultaneous, incoherent driving of the 5s$^2$ $^1$S$_0\leftrightarrow$ 5s5p $^3$P$_1$, 5s5p $^3$P$_1\leftrightarrow$ 5s6s $^3$S$_1$, and 5s5p $^3$P$_2\leftrightarrow$ 5s6s $^3$S$_1$ transitions for 1~ms.  This pumping scheme has the clock state as a unique dark state via the decay of 5s6s~$^3$S$_1$ to the clock state and is in general more robust than coherent driving. However, due to photon recoil, differential trapping, and an unfavorable branching ratio of 5s6s~$^3$S$_1$ to the clock state (requiring many absorption and emission cycles), this process causes significant heating, making it unfavorable as compared to coherent driving. Therefore, we only use this method as a secondary step to transfer atoms left behind by the coherent drive.  

We measure the fidelity of our state transfer by applying a $750~\mu\text{s}$ pulse of intense light resonant with the $^1$S$_0\leftrightarrow$ $^1$P$_1$ transition immediately after state transfer. The large recoil force of this pulse rapidly pushes out atoms in $^1$S$_0$ with a fidelity of $>0.9999$ while leaving atoms in the clock state intact. Upon repumping the clock state back into our imaging cycle and imaging the remaining atoms, we obtain a measure of the fraction of atoms that were successfully transferred to the clock state.  With coherent driving alone, we measure a state transfer fidelity of 0.986(2), while adding incoherent pumping increases this value to 0.998(1). Both of these values are corrected for loss to quantify state transfer in isolation; however, loss also contributes to infidelity of the overall state \textit{preparation}. Taking loss into account, as well as the probability of the atom Raman scattering out of the clock state in the finite time between clock transfer and Rydberg excitation (see Sec.~\ref{sec:spam}), our overall state preparation fidelity with both coherent driving and incoherent pumping is $\mathcal{F}^{\text{SP}}=0.997(1)$.  

\vspace{-3mm}
\section{Auto-ionization and Rydberg state detection fidelity}\label{sec:autoionization}
The auto-ionization beam is resonant with the Sr$^+$ ionic transition $^2$S$_{1/2}$ $\leftrightarrow$ $^2$P$_{3/2}$ at $\lambda_\text{A}=407.6$ nm. The $1/e^2$ beam waist radius is $w_o^\text{A}=16(1)$ $\mu$m with power $P_A=2.8(4)$ mW, from which we estimate a Rabi frequency of $\Omega_A\approx2\pi\times3$~GHz.

To quantify the Rydberg state detection fidelity of our auto-ionization scheme, we compare the observed auto-ionization loss $1/e$ timescale of $\tau_A=35(1)$~ns to the expected lifetime~\cite{Vaillant2012} of $|r\rangle$, which is $\tau_{|r\rangle}\approx80$ $\mu$s. That is, we compute the probability that an atom in the Rydberg state is auto-ionized before it decays away from the Rydberg state. This estimate places an upper bound on the detection fidelity of $|r\rangle$ to be $0.9996(1)$, where the uncertainty is dominated by an assumed uncertainty of $\pm20~\mu\text{s}$ in $\tau_{|r\rangle}$. Note that when the auto-ionization pulse is not applied, there is still a residual detection fidelity of $|r\rangle$ of $0.873(4)$ due to anti-trapping of $|r\rangle$ in the tweezer (this value is smaller than the previously reported $<0.98$ for alkalis in part because the atoms are colder here than in previous work~\cite{Levine2018}).  A lower bound on our detection fidelity is given by the measured $\pi$-pulse fidelity after correcting for errors in preparation and ground state detection, which gives $0.9963(9)$.

We drive the auto-ionizing ion core transition with an intensity that would produce a highly saturated Rabi frequency of $\Omega_A\approx2\pi\times3$~GHz in the bare ion. However, the fast auto-ionization rate~\cite{Cooke1978,Lochead2013} $\Gamma_A > \Omega_A$ of $|r^*\rangle$ actually inhibits the $|r\rangle\leftrightarrow|r^*\rangle$ transition via the continuous quantum Zeno mechanism~\cite{Itano1990,Zhu2014}. In this regime, the effective auto-ionization rate of the transition continues to scale with $\Omega_A^2$ and does not saturate until $\Omega_A \gg \Gamma_A$. This is in qualitative agreement with the fact that our measured auto-ionization loss rate continues to increase with beam intensity. Furthermore, the finite rise time of the acousto-optic modulator (AOM) that we use for switching the auto-ionization beam is a limiting factor in achieving faster auto-ionization. Therefore, detection fidelity can be increased further with higher beam intensity as well as faster beam switching.  

\vspace{-3mm}
\section{State preparation and measurement (SPAM) correction}\label{sec:spam}

At the end of a Rydberg excitation and auto-ionization sequence, we perform state readout by imaging the absence (0) or presence (1) of atoms. We infer the final state of the atom by mapping this binary detection value to the atomic state as $0 \rightarrow |r\rangle$ and $1 \rightarrow |g\rangle$. However, imperfections in state preparation, imaging fidelity, and state-selective readout produce errors in this mapping. State preparation and measurement (SPAM) correction attempts to isolate quantities of the pertinent physics (in this case, Rydberg excitation) from such errors. In particular, we attempt to answer the following question: Assuming an atom is perfectly initialized in the ground state $|g\rangle$, what is the probability that it is in $|r\rangle$ after a certain Rydberg excitation pulse? 

\subsection{Preparation, excitation, and measurement processes}

\begin{figure}[t!]
	\centering
	\includegraphics[width=8cm]{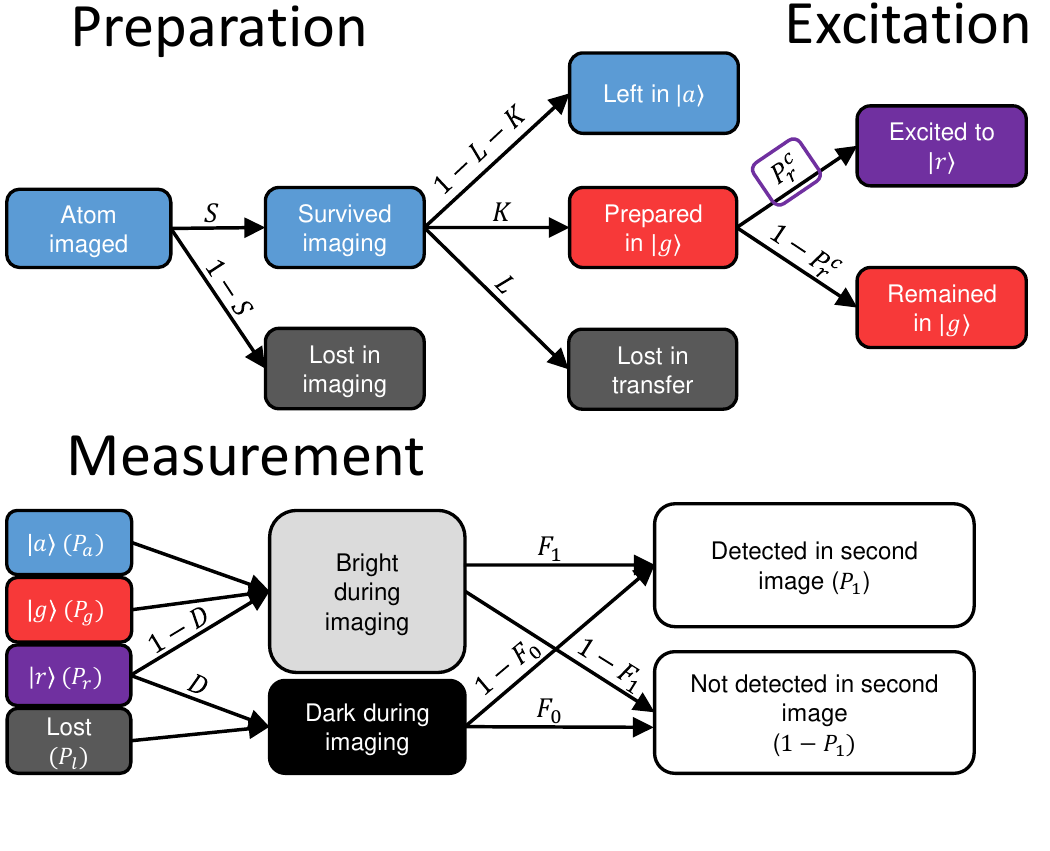}
	\caption{\textbf{Probability tree for single-atom SPAM correction.} Atomic states are color-coded as blue for $|a\rangle$~(absolute ground state), red for $|g\rangle$~(clock state), purple for $|r\rangle$~(Rydberg state), and dark-gray for lost. Quantities above arrows indicate probabilities. The SPAM corrected quantity of interest, $P_r^c$, is highlighted in a purple box. }\vspace{-0.5cm}
	\label{FigFlowchart}
\end{figure}

We begin by assuming that an atom/pair has been registered as present via imaging at the start of the experiment and that it has no detected neighbors within a two tweezer spacing. If an atom/pair does not fulfill this criterion, it is omitted from our data. For the sake of simplicity, we will assume that there are no errors in this initial detection stage. In particular, the combination of high imaging fidelity and high array rearrangement fidelity make errors of this kind exceptionally unlikely. 

Imaging an atom involves a small probability that the atom will be lost, even if it scatters enough photons to be detected. We denote by $S$ the probability that a detected atom survives the first image. After this image, surviving atoms are transferred from the absolute ground state $|a\rangle$ to the clock state $|g\rangle$ (the ground state of our Rydberg qubit) with a probability of successful transfer denoted by $K$. There is a small probability $L$ that during this transfer atoms are lost. The rest, which are not lost but not successfully transferred, remain in $|a\rangle$ with a probability $1-L-K$. The possibilities enumerated up to this point are represented graphically in Extended Data Fig.~\ref{FigFlowchart} under ``Preparation''. 

At this point, atoms that have been successfully prepared in $|g\rangle$ undergo Rydberg excitation. In the single atom case, they end up in the Rydberg state $|r\rangle$ with a probability $P_r^c$. For the two atom case, assuming that both atoms have been successfully prepared, there are four possible states in the two-qubit space, with probabilities given by $P_{rr}^c,P_{rg}^c,P_{gr}^c,P_{gg}^c$. Our ultimate goal will be to solve for these values, which we call ``SPAM-corrected'', indicated here with a superscript $c$. 

In the two atom case, there is the possibility that one atom is successfully prepared while the other is not. In this case, we expect the successfully prepared atom to execute single-atom dynamics. In the case of Rabi oscillations, the Rabi frequency will be reduced by a factor of $\sqrt{2}$. We can thus estimate the Rydberg excitation probability of the prepared atom as $P_{r*}^c \equiv P_r^c|_{\Omega t = \pi} \cos^2(\Omega t/2)$, where $\Omega$ is the single-atom Rabi frequency and $t$ is the pulse length. Of particular interest are the cases $\Omega t = \pi/\sqrt{2}$ and $\Omega t = 2\pi/\sqrt{2}$, corresponding to the two-atom $\pi$ and $2\pi$ pulses, respectively. 

After excitation follows measurement, which involves making Rydberg atoms dark to imaging (i.e., either putting them in a state that scatters no photons or expelling them from the trap) and imaging the remaining bright atoms. In our case, we make Rydberg atoms dark via auto-ionization. We denote by $D$ the probability that a Rydberg atom is successfully made dark to imaging. Furthermore, we denote by $F_0$ the probability of correctly imaging the \textit{absence} of a bright atom (true negative) and by $F_1$ the probability of correctly imaging the \textit{presence} of a bright atom (true positive). $1-F_0$ gives the probability of a false positive, and $1-F_1$ gives the probability of a false negative. 

Let $P_1$ be the probability of an atom being detected as present (bright) at the end of the experiment, and similarly let $P_{00},P_{01},P_{10},P_{11}$ be the corresponding probabilities for atom pairs (with the sum of these being 1). These are the raw, measured values referred to as ``uncorrected'' in the main text and hereafter. 

\subsection{Determining SPAM probabilities}

\begin{table}[b!]
\caption{\textbf{SPAM probabilities.}}
\centering
\begin{tabular}{p{4cm}{c}{c}}
\hline\hline
Probability&Symbol&Value\\
\hline
Imaging true negative & $F_0$ & $0.99997(5)$  \\ 
Imaging true positive & $F_1$ & $0.9988(7)$  \\
Uncorrected survival & $S_0$ & $0.9979(3)$ \\
Corrected survival & $S$ & $0.9991(7)$ \\
Uncorrected $|g\rangle$ transfer & $K_0$ & $0.997(1)$ \\
Raman scattering to $|a\rangle$ & $R$ & $0.00104(1)$ \\
Corrected $|g\rangle$ transfer & $K$ & $0.998(1)$ \\
Loss during $|g\rangle$ transfer & $L$ & $0.0008(8)$ \\
Rydberg state detection & $D$ & $0.9996(1)$ \\
\hline\hline
\end{tabular}
\label{table:SPAM}
\end{table}

We now discuss the determination of the various probabilities discussed. While some of these quantities are directly measurable, some must be estimated from measurements that themselves need SPAM correction. All probabilities entering into SPAM correction calculations are summarized in Extended Data Table~\ref{table:SPAM}.

We determine $F_0$ and $F_1$ by analyzing the histogram of detected photons from a typical set of images, similarly to the method described in Ref.~\cite{Cooper2018a}. The histograms have a zero- and one-atom peak, and we determine false positives and false negatives by the area of these peaks that extends beyond the binary detection threshold. Loss during imaging that leads to false negatives is also taken into account in $F_1$~\cite{Cooper2018a}. Error bars are given by the standard deviation across the array. 

We determine $S$ by taking two consecutive images. We measure the value $S_0$, defined as the probability of detecting an atom in the second image conditional on its detection in the first. Obtaining the true value of $S$ from $S_0$ requires correcting for false positives and false negatives in the second image (where we assume false positives in the first image are negligible). One can write $S_0$ as the sum of atoms that survived and were correctly positively identified and that did not survive and were incorrectly positively identified. Solving for $S$ gives:

\begin{align}
S&=\frac{S_0+F_0-1}{F_0+F_1-1}
\end{align}

By a similar procedure, we determine $K$ from a value $K_0$ measured by performing state transfer, using a ground-state push-out pulse as described in Sec.~\ref{SecStatePrep}, repumping to the ground state, and measuring the probability of detecting an atom in a subsequent image. To obtain the true $K$, we correct $K_0$ for imaging errors as well as survival probability after imaging. We furthermore modify $K$ with the probability $R$ that a successfully transferred atom goes back to $|a\rangle$ due to trap Raman scattering in the time delay between state transfer and Rydberg excitation. We estimate $R = 0.00104(1)$ by a measure of the lifetime in the clock state at our tweezer depth~\cite{Covey2019a}. We obtain:

\begin{align}
K&=\frac{K_0+F_0-1}{S_0+F_0-1}(1-R).
\end{align}

We note that the total clock state preparation fidelity, an important quantity on its own, can be expressed as $\mathcal{F}^{\text{SP}}=SK=0.997(1)$. To measure the transfer loss probability $L$, we perform state transfer without a push-out pulse, then repump atoms to the ground state and measure how many were lost (again correcting for imaging loss and imaging errors). 

Finally, we determine $D$ by comparing the measured auto-ionization timescale to an estimate of the Rydberg lifetime, as described in Sec.~\ref{sec:autoionization}. We assume all decay from the Rydberg state is into bright states and therefore leads to detection errors, which is physically motivated by the large branching ratio of our Rydberg state to the 5s5p$~^3$P$_J$ manifold, whose states are repumped into our imaging cycle. We neglect other processes that may make a Rydberg atom go dark, such as anti-trapping or decay into dark states, as these are expected to have a much longer timescale. 

\vspace{-3mm}
\subsection{Correcting the single-atom excitation probabilities}

\begin{table}[b!]
\caption{\textbf{Possible states for a single atom.} Note that the sum of these populations equals unity.}
\centering
\begin{tabular}{p{4cm}{c}{c}}
\hline\hline
State&Symbol&Value\\
\hline
Lost & $p_l$ & $(1-S)+SL$ \\
$|a\rangle$ (absolute ground state) & $p_a$ & $S(1-L-K)$ \\
$|g\rangle$ (clock state) & $p_g$ & $SK(1-P^{c}_{r})$ \\
$|r\rangle$ (Rydberg state) & $p_r$ & $SKP^{c}_{r}$ \\
\hline\hline
\end{tabular}
\label{table:States1}
\end{table}

We are now ready to solve for $P_r^c$ in terms of the uncorrected value $P_1$ and the various SPAM probabilities. For clarity, it will be convenient to define variables for the populations of the four possible single-atom states that an atom can be in at the end of Rydberg excitation: lost, $|a\rangle$, $|g\rangle$, and $|r\rangle$. We will call these populations $p_l$, $p_a$, $p_g$ and $p_r$, respectively, with their values determined by the probability tree in Extended Data Fig.~\ref{FigFlowchart} and summarized in Extended Data Table~\ref{table:States1}. 

We can write $P_1$ as a sum of true positive identifications of bright states plus false positive identification of dark states (see ``Measurement'' in Extended Data Fig.~\ref{FigFlowchart}). In terms of the values defined so far, we have:

\begin{align}
P_1=(p_a+p_g+p_r(1-D))F_1+(p_l+p_rD)(1-F_0).
\end{align}

Substituting in the full expression for the populations from Extended Data Table~\ref{table:States1} and solving for $P_r^c$, we obtain:

\begin{align}
P_r^c = \frac{SF_1+(1-S)(1-F_0)-LS(F_0+F_1-1)-P_1}{KSD(F_0+F_1-1)}.
\end{align}

For the single-atom short-time Rabi oscillations reported in Table I of the main text, we observe the bare values of $P_1(\pi)=0.0049(9)$ and $P_1(2\pi)=0.9951(9)$, yielding SPAM-corrected pulse fidelities of $\mathcal{F}^{\text{SPAM}}(\pi)=P_r^c(\pi)=0.9967(9)$ and $\mathcal{F}^{\text{SPAM}}(2\pi)=1-P_r^c(2\pi)=0.998(1)$, respectively.

\vspace{-3mm}
\subsection{Correcting the two-atom excitation probabilities}

\begin{table}[b!]
\caption{\textbf{Possible states for two atoms.} Note that the sum of these populations equals unity. Terms inside $\{\}$ have an implied symmetric partner, e.g. $p_{al}\equiv p_{la}$.}
\centering
\begin{tabular}{p{3cm}{c}{c}{c}{c}}
\hline\hline
States&Symbol&Value\\
\hline
Lost, Lost & $p_{ll}$ & $((1-S)+SL)^2$ \\

\{Lost, $|a\rangle$\} & $p_{la}$ & $((1-S)+SL)S(1-L-K)$ \\

\{Lost, $|g\rangle$\} & $p_{lg}$ & $((1-S)+SL)SK(1-P_{r*}^c)$ \\

\{Lost, $|r\rangle$\} & $p_{lr}$ & $((1-S)+SL)SKP_{r*}^c$  \\

$|aa\rangle$ & $p_{aa}$ & $S^2(1-L-K)^2$  \\

\{$|ag\rangle$\} & $p_{ag}$ & $S(1-L-K)SK(1-P_{r*}^c)$  \\

\{$|ar\rangle$\} & $p_{ar}$ & $S(1-L-K)SKP_{r*}^c$  \\

$|gg\rangle$ & $p_{gg}$ & $S^2K^2(1-P^{c}_{rg}-P^{c}_{gr}-P^{c}_{rr})$  \\

$|gr\rangle$ & $p_{gr}$ & $S^2K^2P^{c}_{gr}$  \\

$|rg\rangle$ & $p_{rg}$ & $S^2K^2P^{c}_{rg}$  \\

$|rr\rangle$ & $p_{rr}$ & $S^2K^2P^{c}_{rr}$  \\
\hline\hline
\end{tabular}
\label{table:States2}
\end{table}

For the two-atom case, there are 16 possible states for an atom pair. Similarly to Extended Data Table~\ref{table:States1}, we can write populations of each of these states in terms of the survival and transfer fidelities in Extended Data Table~\ref{table:SPAM}, as shown in Extended Data Table~\ref{table:States2}.

We now write the experimentally measured quantities $P_{10}, P_{00}$, and $P_{11}$ in terms of the values in Extended Data Tables~\ref{table:SPAM} and~\ref{table:States2}. For notational simplicity we define $\bar{F_0}\equiv(1-F_0)$, and similarly for $F_1$ and $D$:
\begin{eqnarray}
P_{10}&&=p_{ll}(\bar{F_0}F_0)\nonumber\\
	  &&+p_{la}(\bar{F_0}\bar{F_1})\nonumber\\
	  &&+p_{al}(F_1F_0)\nonumber\\
	  &&+p_{lg}(\bar{F_0}\bar{F_1})\nonumber\\
	  &&+p_{gl}(F_1F_0)\nonumber\\
	  &&+p_{lr}(\bar{F_0}F_0D+\bar{F_0}\bar{D}\bar{F_1})\nonumber\\
	  &&+p_{rl}(\bar{F_0}DF_0+F_1\bar{D}F_0)\nonumber\\
	  &&+p_{aa}(F_1\bar{F_1})\nonumber\\
	  &&+p_{ag}(F_1\bar{F_1})\nonumber\\
	  &&+p_{ga}(F_1\bar{F_1})\nonumber\\
	  &&+p_{ar}(F_1 D F_0+F_1\bar{D}\bar{F_1})\nonumber\\
	  &&+p_{ra}(F_1\bar{D}\bar{F_1}+\bar{F_0}D\bar{F_1})\nonumber\\
	  &&+p_{gg}(F_1\bar{F_1})\nonumber\\
	  &&+p_{gr}(F_1 D F_0+F_1\bar{D}\bar{F_1})\nonumber\\
	  &&+p_{rg}(F_1\bar{D}\bar{F_1}+\bar{F_0}D\bar{F_1})\nonumber\\
	  &&+p_{rr}(F_1\bar{D}F_0 D+\bar{F_0}D\bar{F_1}\bar{D}+\bar{F_0}F_0 D^2+F_1\bar{F_1}\bar{D}^2),\nonumber\\
\end{eqnarray}

\begin{eqnarray}
P_{00}&&=p_{ll}(F^2_0)\nonumber\\
	  &&+p_{la}(F_0\bar{F_1})\nonumber\\
	  &&+p_{al}(\bar{F_1}F_0)\nonumber\\
	  &&+p_{lg}(F_0\bar{F_1})\nonumber\\
	  &&+p_{gl}(\bar{F_1}F_0)\nonumber\\
	  &&+p_{lr}(F^2_0 D+F_0\bar{F_1}\bar{D})\nonumber\\
	  &&+p_{rl}(F^2_0 D +\bar{F_1}\bar{D}F_0)\nonumber\\
	  &&+p_{aa}(\bar{F_1}^2)\nonumber\\
	  &&+p_{ag}(\bar{F_1}^2)\nonumber\\
	  &&+p_{ga}(\bar{F_1}^2)\nonumber\\
	  &&+p_{ar}(\bar{F_1}F_0 D+\bar{F_1}^2\bar{D})\nonumber\\
	  &&+p_{ra}(\bar{F_1}^2\bar{D}+F_0 D\bar{F_1})\nonumber\\
	  &&+p_{gg}(\bar{F_1}^2)\nonumber\\
	  &&+p_{gr}(\bar{F_1}F_0 D+\bar{F_1}^2\bar{D})\nonumber\\
	  &&+p_{rg}(\bar{F_1}^2\bar{D}+F_0 D\bar{F_1})\nonumber\\
	  &&+p_{rr}(\bar{F_1}^2\bar{D}^2+F_0 D\bar{F_1}\bar{D}+F^2_0 D^2+\bar{F_1}\bar{D}F_0 D),\nonumber\\
\end{eqnarray}

\begin{eqnarray}
P_{11}&&=p_{ll}(\bar{F_0}^2)\nonumber\\
	  &&+p_{la}(\bar{F_0}F_1)\nonumber\\
	  &&+p_{al}(F_1\bar{F_0})\nonumber\\
	  &&+p_{lg}(\bar{F_0}F_1)\nonumber\\
	  &&+p_{gl}(F_1\bar{F_0})\nonumber\\
	  &&+p_{lr}(\bar{F_0}^2D+\bar{F_0}F_1\bar{D})\nonumber\\
	  &&+p_{rl}(\bar{F_0}^2D+F_1\bar{D}\bar{F_0})\nonumber\\
	  &&+p_{aa}(F^2_1)\nonumber\\
	  &&+p_{ag}(F^2_1)\nonumber\\
	  &&+p_{ga}(F^2_1)\nonumber\\
	  &&+p_{ar}(F_1 \bar{F_0} D+F^2_1\bar{D})\nonumber\\
	  &&+p_{ra}(F^2_1\bar{D}+\bar{F_0} D F_1)\nonumber\\
	  &&+p_{gg}(F^2_1)\nonumber\\
	  &&+p_{gr}(F_1 \bar{F_0} D+F^2_1\bar{D})\nonumber\\
	  &&+p_{rg}(F^2_1\bar{D}+\bar{F_0} D F_1)\nonumber\\
	  &&+p_{rr}(F^2_1\bar{D}+\bar{F_0}D F_1\bar{D}+\bar{F_0}^2D^2\bar{D}+F_1\bar{D}\bar{F_0} D).\nonumber\\
\end{eqnarray}

Note that $P_{01}=1-P_{10}-P_{00}-P_{11}$. Thus, with the three above equations, we can solve for $P^{c}_{gg}$, $P^{c}_{rg}$, $P^{c}_{gr}$, and $P^{c}_{rr}$. The full expressions for these solutions are cumbersome and not shown. The experimentally measured values $P_{00}$, $P_{10}$, $P_{01}$ and $P_{11}$ are reported in Extended Data Table~\ref{table:MeasuredTwo}. 

\begin{table}[b]
\caption{\textbf{Experimentally measured two-atom values.} Uncorrected values used to calculate $P^{c}_{gg}$, $P^{c}_{rg}$, $P^{c}_{gr}$, and $P^{c}_{rr}$ at both the $\pi$- and $2\pi$-times. The `T' superscript indicates the values for which the traps were on. We report the values of $P_{10}$ and $P_{01}$ in symmetrized and antisymmetrized form, where $P_{\{10\}}=P_{10}+P_{01}$ and $P_{[10]}=P_{10}-P_{01}$}
\centering
\begin{tabular}{p{4cm}{c}}
\hline\hline
Variable&Value\\
\hline
$P_{\{10\}}(\pi)$ & 0.992(2)\\
$P_{[10]}(\pi)$ & 0.01(1)\\
$P_{00}(\pi)$ & 0.0032(7)\\

$P_{11}(2\pi)$ & 0.989(2)\\
$P_{[10]}(2\pi)$ & 0.004(2)\\
$P_{00}(2\pi)$ & 0.0036(7)\\

$P^{T}_{\{10\}}(\pi)$ & 0.992(2)\\
$P^{T}_{[10]}(\pi)$ & 0.004(10)\\
$P^{T}_{00}(\pi)$ & 0.0032(7)\\

$P^{T}_{11}(2\pi)$ & 0.985(2)\\
$P^{T}_{[10]}(2\pi)$ & -0.003(2)\\
$P^{T}_{00}(2\pi)$ & 0.0030(6)\\
\hline\hline
\end{tabular}
\label{table:MeasuredTwo}
\end{table}

\vspace{-3mm}
\section{Bell state fidelity}\label{sec:bell}
\subsection{Bounding the Bell state fidelity}
Characterizing the state of a quantum system is of fundamental importance in quantum information science. Canonical tomographic methods addressing this task require a measurement of a complete basis set of operators. Such measurements are often expensive or not accessible. More economic approaches can be employed to assess the overlap with a given target state. For example the overlap of a two-qubit state with a Bell state is routinely determined by measuring the populations in the four computational basis states (yielding the diagonal elements of the density operator), in addition with a measurement that probes off-diagonal elements via parity oscillations~\cite{Liebfried2005,Levine2018}. To access the latter it is however necessary to perform individual, local operations on the qubits. Here, we present a bound on the Bell state fidelity that can be accessed with only global control and measurements in the computational basis and elaborate on the underlying assumptions. 

Specifically, we are interested in the overlap $\mathcal{F}$ of the experimentally created state $\rho$ with a Bell state of the form $\ket{W_{\phi}}=\tfrac{1}{\sqrt{2}}(\ket{gr}+e^{i\phi}\ket{rg})$. This is defined as
\begin{align}
\mathcal{F}=\max_\phi~\bra{W_{\phi}}\rho\ket{W_{\phi}}=\frac{1}{2}\big(\rho_{gr,gr}+\rho_{rg,rg}+2|\rho_{gr,rg}|\big).
\end{align}
Here we denote matrix elements of a density operator $\rho$ in the two-atom basis by $\rho_{i,j}=\bra{i}\rho\ket{j}$, with $i,j \in \{gg,gr,rg,rr\}$. Clearly, a measurement of $\mathcal{F}$ requires access to the populations in the ground and Rydberg states $\rho_{i,i}$ as well as some of the coherences $\rho_{i,j}$ with $i\neq j$. While populations are direct observables (in particular, we identify $\rho_{i,i}$ with our measured values $P_i$ or their SPAM corrected counterparts $P_i^c$), coherences are not. We can however bound the fidelity $\mathcal{F}$ from below via a bound on $|\rho_{gr,rg}|$. Namely, it can be shown via Cauchy's inequality $|\rho_{a,b}|^2\leq \rho_{a,a} \rho_{b,b}$ and the normalization of states $\sum_i \rho_{i,i}=1$ that 
\begin{eqnarray}\label{diagbound}
&&|\rho_{gr,rg}|^2 \geq \frac{1}{2}(\tr{\rho^2}-1) + \rho_{gr,gr} \rho_{rg,rg} \nonumber\\
\end{eqnarray}
where $\tr{\rho^2} = \sum_{i,j} |\rho_{i,j}|^2$ is the purity. 

Evaluating the bound given by equation Eq.~\eqref{diagbound} requires access to the purity (or a lower bound thereof). One can bound the purity from below by the populations in the ground and Rydberg states as 
\begin{align}\label{purity}
\tr{\rho^2}\geq \sum_{i}(\rho_{i,i})^2.
\end{align}
In general Eq.~\eqref{purity} is a very weak bound. In particular, it does not distinguish between a pure Bell state $\ket{\psi_{\phi}}$ and the incoherent mixture of the two states $\ket{gr}$ and $\ket{rg}$. 
However, if the state $\rho$ is close to one of the four atomic basis states (as is the case at the $2\pi$ time of the Rabi evolution), the bound Eq.~\eqref{purity} becomes tight. 
This fact allows us to obtain a lower bound for the purity of the Bell state in the experiment as follows. The Bell state in our protocol is generated by evolving the state $\ket{gg}$ for a time $T=\pi/\tilde{\Omega}_R$ in the Rydberg-blockade regime. Note that the same evolution should lead to a return to the initial state at time $2T$ in the ideal case. Under the \textit{assumption} that a coupling to the environment decreases the purity of the quantum system (see further exploration of this assumption in the following subsection), we can bound the purity of the state at time $T$ by the purity of the state at time $2T$, which in turn can be bounded by measurements of the atomic populations at time $2T$ via Eq.~\eqref{purity}:
\begin{align}
\tr{\rho(T)^2}\geq \tr{\rho(2T)^2}\geq  \sum_{i}\rho_{i,i}(2T)^2.
\end{align}
Using this estimated bound on the purity leads to a lower bound on the Bell state fidelity $\mathcal{F}$ at time $T$ solely in terms of the populations in the ground and Rydberg states at times $T$ and $2T$:
\begin{eqnarray}\label{bound}
&&\mathcal{F}(T)\geq \frac{1}{2}\bigg(\rho_{gr,gr}(T)+\rho_{rg,rg}(T)\nonumber\\
&&+2\sqrt{\max\lr{0,(\textstyle\sum_{i}\rho_{i,i}(2T)^2-1)/{2}+\rho_{gr,gr}(T)\rho_{rg,rg}(T)}}\bigg).\nonumber\\
\end{eqnarray}

\subsection{Bounding an increase in purity due to spontaneous decay}
Although we make the assumption that the purity of our state does not increase between times $T$ and $2T$ and assert that this assumption is reasonable, we recognize the hypothetical possibility that dissipative processes such as spontaneous emission can in principle increase the purity of quantum states. We note an increase of purity with time typically occurs only in specially engineered situations (as in optical pumping schemes), and we have no reason to believe such mechanisms are active in our system. In fact, reasonable numerical models of potential decoherence mechanisms are all consistent with a decrease of the purity. Nevertheless, we now analyze how strongly our assumption of purity decrease could potentially be violated given the spontaneous emission rate of our Rydberg state and show that the corresponding decrease of the inferred Bell state fidelity is well within our confidence interval.

We assume that the system can be modeled by a Markovian Master equation of the form:
\begin{eqnarray}\label{markovian}
\dot \rho=\mc{L}\rho=-i[H,\rho]&&+\sum_{\mu} \gamma_\mu (c_{\mu}\rho c_\mu^\dag-\frac{1}{2}\{c_\mu^\dag c_\mu,\rho\})\nonumber\\
&&+\sum_{\mu}\bar\gamma_\mu (h_{\mu}\rho h_\mu-\frac{1}{2}\{h_\mu h_\mu,\rho\})\nonumber\\
\end{eqnarray}
Here we explicitly distinguish incoherent terms generated by Hermitian jump operators ($h_\mu=h_\mu^\dag$, e.g. dephasing), and non-Hermitian jump operators ($c_\mu$, e.g. spontaneous emission). We find
\begin{align}\label{me1}
\frac{d}{dt}\tr{\rho^2}=2\tr{\rho(\mc{L}\rho)}\leq2\sum_{\mu} \gamma_\mu \tr{\rho c_{\mu}\rho c_\mu^\dag-c_\mu^\dag c_\mu\rho^2}
\end{align}
which simply reflects the fact that the purity of the quantum state can not increase due to the coherent part of the evolution or due to any incoherent part of the evolution that is generated by Hermitian jump operators (dephasing). Thus the coherent part of the evolution does not affect the bound we obtain in the end. Eq.~\eqref{me1} can be obtained from Eq.~\eqref{markovian} by noting that $\tr{\rho[H,\rho])}=\tr{\rho H\rho-\rho^2H}=0$ and $\tr{\rho[h_{\mu},[\rho, h_\mu]]}=-\tr{[h_{\mu},\rho][\rho, h_\mu]}=-\tr{([\rho, h_\mu])^\dag [\rho, h_\mu]}\leq 0$, which gives Eq.~\eqref{me1}.

Now let us assume that the non-Hermitian jump operators correspond to decay from the Rydberg state $\ket{r}$ into some set of states $\{\ket{f}|f=1,2,\dots n\}$ that also include the ground state $\ket{g}\equiv \ket{1}$. The following argument works for arbitrary $n\geq 1$. Since we have two atoms we have $2n$ non-Hermitian jump operators $c_{f}^{(a)}=\ket{f}_a\!\bra{r}$, where $a=1,2$ labels the atoms.
With this model we have (denoting the reduced state of atom $a$ by $\rho^{(a)}$):
\begin{eqnarray}
\frac{d}{dt}\tr{\rho^2}&&\leq2\sum_{f,a} \Gamma_f \tr{\rho c_{f}^{(a)}\rho {c_{f}^{(a)}}^\dag-{c_{f}^{(a)}}^\dag c_{f}^{(a)}\rho^2}\nonumber\\
&&=2\sum_{f,a} \Gamma_f(\rho^{(a)}_{f,f}\rho^{(a)}_{r,r}- \rho^{(a)}_{r,r}\rho^{(a)}_{r,r}-\sum_{e\neq r} \rho^{(a)}_{r,e}\rho^{(a)}_{e,r})\nonumber\\
&&\leq2\sum_{f,a} \Gamma_f (\rho^{(a)}_{f,f}\rho^{(a)}_{r,r}- \rho^{(a)}_{r,r}\rho^{(a)}_{r,r})\nonumber\\
\end{eqnarray}
where $\Gamma_f$ is the single-atom decay rate from $|r\rangle$ to $|f\rangle$. Note that $\rho^{(a)}_{f,f}\rho^{(a)}_{r,r}- \rho^{(a)}_{r,r}\rho^{(a)}_{r,r}\leq(1-\rho^{(a)}_{r,r})\rho^{(a)}_{r,r}- \rho^{(a)}_{r,r}\rho^{(a)}_{r,r}\leq1/8$. This gives the final result
\begin{align}
\frac{d}{dt}\tr{\rho^2}&\leq\frac{1}{2}\sum_f\Gamma_f =\frac{1}{2}\Gamma
\end{align}
That is, the rate at which the purity increases is upper bounded by half the rate at which a single atom in the Rydberg state decays into other states by spontaneous emission. Over a time interval of length $T$ the 2-atom purity can thus not increase by more than $T\Gamma/2$. 

Using our blockaded $\pi$-time for $T$ and Rydberg state decay rate for $\Gamma$, we evaluate this bound on the purity increase to be $3.2\times10^{-4}$. This would lead to a decrease in our bound on the Bell state fidelity by $1.6\times10^{-4}$ for both the cases of tweezers off and tweezers on, which is significantly smaller than our quoted error for these values. 

\section{Rydberg laser system}
The Rydberg laser system is based on a Toptica laser, in which an extended cavity diode laser (ECDL) at $\lambda_\text{IR}=1266.6$ nm seeds a tapered amplifier (TA) with output power up to $\approx2$ W, which is then frequency doubled via second harmonic generation (SHG) in a bowtie cavity to obtain up to $\approx1$ W at $\lambda_\text{Red}=633.3$ nm, which is then frequency doubled in a second bowtie cavity to obtain fourth harmonic generation (FHG) with up to $\approx0.4$ W at $\lambda_\text{UV}=316.6$ nm. The fundamental laser at $\lambda_\text{IR}=1266.6$ nm is stabilized to an ultralow expansion (ULE) cavity system (Stable Laser Systems) of length 10 cm with finesse of $\approx14000$ and line width (full width at half maximum) of $\approx110$ kHz. The finesse was measured by performing cavity ringdown spectroscopy~\cite{Anderson1984}. We currently do not filter the fundamental laser with the cavity~\cite{Levine2018}, but we are prepared to implement this approach. Further discussion on the laser frequency stability can be found in Appendix~\ref{SecDecoherence}.

We use a beam power of $P_R=28.1(4)$ mW, measured immediately before it enters the vacuum cell (through 4~mm of uncoated quartz). The geometric mean $1/e^2$ waist radius of the beam at the position of the atoms is $\overline{w}_0^R=18(1)$ $\mu$m. These conditions correspond to the Rabi frequency used throughout the text of $\Omega_R\approx2\pi\times6-7$ MHz. The maximum power we can achieve is $\approx110$ mW, for which we observe a Rabi frequency of $\approx2\pi\times13$ MHz. The Rydberg pulses are derived from an AOM, which limits the rise and fall time to $\approx40$ ns.  When driving with a Rabi frequency whose $\pi$-pulse approaches this timescale ($\Omega_R\approx2\pi\times13$ MHz), we observe an asymmetric reduction in Rabi signal contrast by $\approx1-2\%$ at the multiples of $2\pi$, unlike conventional detuned Rabi oscillations where the contrast reduction occurs at the odd multiples of $\pi$. We attribute to early-time dynamics during the AOM switching. We do not use an optical fiber, so there is limited spatial -- and thus spectral -- filtering between the AOM and the atoms. Accordingly, we intentionally work with $\Omega_R\approx2\pi\times6-7$ MHz such that the $\pi$-pulse time is sufficiently slow compared to the AOM rise and fall times. However, when operating at $\Omega_R\approx2\pi\times13$ MHz we observe long-time coherence similar to, or slightly better than, the reported values in the main text for $\Omega_R\approx2\pi\times6-7$ MHz. Measured results under all conditions are consistent with the numerical analysis summarized in Extended Data Fig.~\ref{FigRabi} below.

\vspace{-3mm}
\section{Rydberg decoherence mechanisms}\label{SecDecoherence}
For a non-interacting case where Rydberg atoms in a tweezer array are well separated, the Hamiltonian $H$ driving Rabi oscillations is
\begin{align}\label{Ham}
	H = \sum_{i=1}^N \Omega_{R,i} S_i^x +  \Delta_i S_i^z,
\end{align}
where $\Omega_{R,i}$ and $\Delta_i$ are the Rabi frequency and the detuning for the atom at site $i$, $S^{\mu}$ are the spin-1/2 operators with $\mu=x,y,z$, and $N$ is the total number of atoms. Variations in the Rabi frequency and detuning, manifesting either as non-uniformity across the tweezer array (e.g. from non-uniform beam alignment) or as random noise, lead to a decay in the array-averaged Rabi signal. In our system, we measure a $1/e$ decay time of $\approx$7~$\mu s$ at a Rabi frequency of 6 MHz (see Fig. 3a of the main text). In this section, we present a model of decoherence mechanisms that accounts for our observed decay. 

As a preliminary, we begin by confirming that the spatial variation of Rabi frequency across different tweezers is less than 0.2\%, and that no variation of detuning across the array is observed. We conclude that non-uniformity is not a dominant contributor to our observed Rabi decay. 

Therefore, we focus here on three factors that induce random noise in the Rabi frequency and detuning: atomic motion, laser phase noise, and laser intensity noise. We perform Monte Carlo-based simulations~\cite{deLeseleuc2018} that take into account these noise sources as well as the finite lifetime $\approx$80$~\mu$s of the $n=61$ Rydberg state due to spontaneous emission. In the following subsections, we discuss relative contributions from these noise sources.

\vspace{-3mm}
\subsection{Atomic motion}
An atom with a nonzero momentum shows a Doppler shift relative to the bare resonance frequency. At the beginning of Rabi interrogation, the momentum distribution, and thus the distribution of Doppler shifts, follows that of an atom in a trap. More specifically, for an atom at temperature $T$ trapped in a harmonic potential with the radial trap frequency $\omega_r$, the Doppler shift distribution can be modeled as a normal distribution with the standard deviation $\Delta_T$:
\begin{align}\label{Doppler}
	\Delta_T = \frac{k_L}{m} \sqrt{\frac{\hbar m\omega_r}{2 \tanh(\hbar\omega_r/ 2k_B T)}},
\end{align}
where $m$ is the mass of $^{88}$Sr and $k_L$ is the wavevector of the Rydberg excitation light.

The radial temperature of our atomic array (along the axis of propagation of the Rydberg beam) is measured via sideband spectroscopy on the clock transition \cite{Madjarov2019} to be $T_r \approx 2.5~\mu\text{K}$ at a radial trap frequency of $\omega_r\approx2\pi\times78$~kHz. We adiabatically ramp down the trap by a factor of 10 before Rydberg interrogation, thereby reducing the temperature and the trap frequency by a factor of $\sqrt{10}$ (which we also confirm via further sideband spectroscopy). Using Eq.~\ref{Doppler}, we estimate the Doppler broadening to be $\Delta_T\approx2\pi\times30$~kHz. At a Rabi frequency of $\Omega_R\approx2\pi\times6$ MHz, the expected Doppler decoherence timescale is $\tau  \sim \Omega_R/\Delta_T^2$ $\approx$10~ms, which is three orders of magnitudes longer than the measured value $\approx$7~$\mu$s. This implies that motional effects are negligible in the Rabi decoherence dynamics.

\begin{figure}[t!]
	\centering
	\includegraphics[width=7cm]{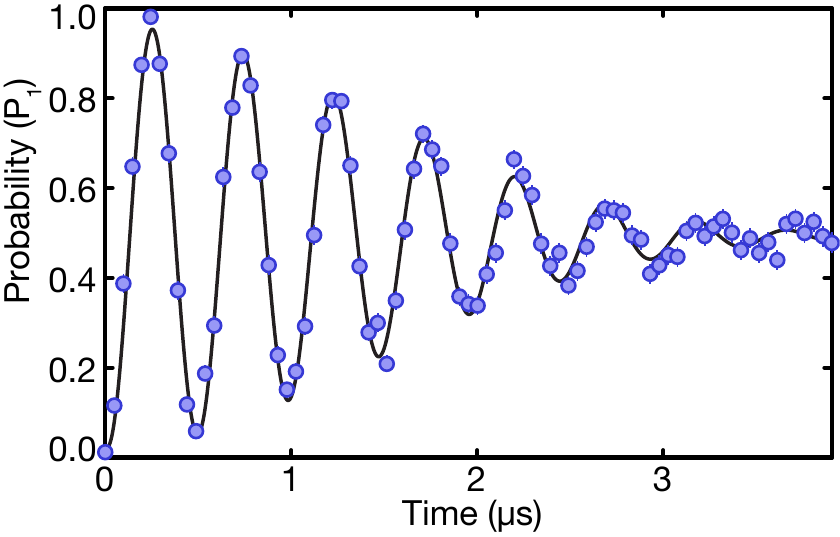}
	\caption{\textbf{Ramsey interferometry.} We use a detuning of 2 MHz between the two pulses to show oscillations with a characteristic $1/e$ decay time $\tau_\text{Ramsey}\approx2~\mu\text{s}$. A sine-modulated Gaussian decay is used for the fit (solid line). Data is uncorrected and averaged over $\approx40$ experimental cycles per timestep and over an array of approximately 14 atoms. Error bars indicate a $1\sigma$ binomial confidence interval.}\vspace{-0.5cm}
	\label{FigRamsey}
\end{figure}

\vspace{-3mm}
\subsection{Laser phase noise}
Phase noise manifests as random temporal fluctuation of the detuning $\Delta$ in the Hamiltonian in Eq.~\ref{Ham}. Since the frequency of the Rydberg laser is stabilized to a ULE reference cavity via the Pound-Drever-Hall (PDH) method, we use an in-loop PDH error signal derived from the cavity reflection to extract a phase noise spectrum (see Ref.~\cite{deLeseleuc2018} for the detailed procedures of phase noise extraction). The obtained noise power spectral density, predicting a RMS frequency deviation of $\approx$0.6~MHz after fourth-harmonic generation, allows us to generate random time-varying detuning profiles that are fed into our Monte Carlo simulations to extract a predicted decay time. Note that while the estimated laser linewidth is $\sim1-10$ kHz, phase noise from the servo bumps centered at $\nu_\text{SB}\approx0.6$ MHz is highly relevant since $\Omega_R>\nu_\text{SB}$, and in fact dominates the RMS. 

Since the cavity filters phase noise beyond its linewidth, this noise is suppressed on the measured PDH signal as compared to the actual noise of the laser light that we use for Rydberg interrogation. We therefore correct our measured phase noise spectrum with a cavity roll-off factor~\cite{Nagourney2014} obtained from the cavity linewidth and finesse, which results in an increase in noise as compared to the uncorrected measured spectrum. However, we can also use the uncorrected spectrum to predict the phase noise we would have if we used the filtered cavity light to generate our Rydberg light via a technique described in Ref.~\cite{Levine2018}. The results in Extended Data Fig.~\ref{FigRabi} show simulated results both with and without cavity filtering. 

Our simulations (without cavity filtering, as in our current setup) predict a Ramsey decay time of $\approx2~\mu\text{s}$ with a Gaussian envelope, which is consistent with our experimental observation. In principle, Doppler broadening $\Delta_T$ could also lead to dephasing in Ramsey signals; however, the corresponding $1/e$ decay time is expected to be $\tau_\text{Ramsey} = \sqrt{2}/\Delta_T = 7.5~\mu s$, longer than the observed 2~$\mu$s, suggesting that laser phase noise is dominant over motional effects in our Ramsey signal.

\vspace{-3mm}
\subsection{Laser intensity noise}
Our intensity noise predominantly originates directly from the Rydberg laser. This intensity noise is composed of both high-frequency fluctuations compared to the pulse length, and lower frequency (effectively shot-to-shot) fluctuations. Using a UV avalanche photodetector (APD130A2, Thorlabs), we measure that the intensity pulse areas between different experimental trials are normally distributed with fractional standard deviation $\sigma_\text{RMS} \sim 1/\sqrt{L}$, where $L$ is the pulse duration, saturating to $0.8\%$ when $L > 1$ $\mu$s. Note that the pulses are too fast to stabilize with an AOM during interrogation, and that we employ a sample-and-hold method.

\begin{figure}[t!]
	\centering
	\includegraphics[width=7cm]{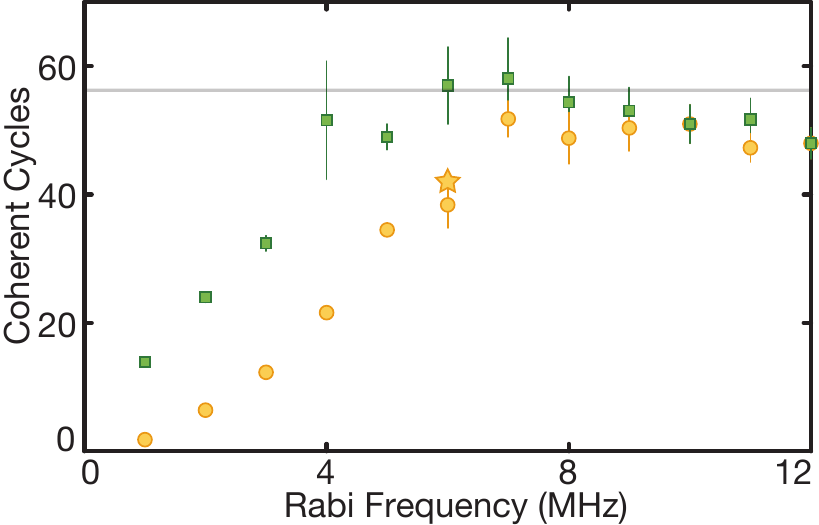}
	\caption{\textbf{Simulated and measured $1/e$ coherence vs Rabi frequency.} The star represents the measured data shown in Fig. 3a, and the circle and square points represent numerical modeling with measured laser phase and intensity noise profiles. The yellow circles show the case when cavity phase noise filtering is \textit{not} performed (as in this work), and the green squares show the case where cavity phase noise filtering \textit{is} performed. The horizontal gray line shows the upper limit due to measured intensity noise fluctuations with RMS deviation of $0.8\%$ (see Eq.~\eqref{NRabi}). Error bars indicate a $1\sigma$ confidence interval.}\vspace{-0.5cm}
	\label{FigRabi}
\end{figure}

In the presence of only intensity noise following a normal distribution with fractional standard deviation $\sigma_\text{RMS}$, one can closely approximate the noise in the Rabi frequency to also be normally distributed and derive an analytical expression for a $1/e$ Rabi decay time as $\tau_\text{Rabi}  = 2\sqrt{2}/(\Omega_R \sigma_\text{RMS})$ where $\Omega_R$ is the nominal, noise-free Rabi frequency. In the intensity noise limited regime, we thus expect a Rabi lifetime $\mathcal{N}_\text{Rabi}$ (in oscillation cycles) to be Rabi frequency-independent (see the line in Extended Data Fig.~\ref{FigRabi}):
\begin{align}\label{NRabi}
	\mathcal{N}_\text{Rabi} = \frac{\Omega_R \tau_\text{Rabi}}{2\pi} = \frac{\sqrt{2}}{\pi\sigma_\text{RMS}}.
\end{align}

\subsection{Summary}
Including all the discussed noise sources (atomic motion, phase noise, intensity noise) as well the finite state lifetime and a Rydberg probe-induced light shift (discussed in a subsequent section), we calculate $\mathcal{N}_\text{Rabi}$ as a function of drive frequency, as shown in Extended Data Fig.~\ref{FigRabi}. We find that the simulated Rabi oscillation agrees well with the experimental result at a Rabi frequency of 6 MHz. While the Rabi lifetime improves with increasing Rabi frequency, it becomes saturated to $\mathcal{N}_\text{Rabi} \approx 56$ at high Rabi frequencies due to intensity noise fluctuations. Interestingly, we note that there is a crossover between a phase noise-limited regime at low Rabi frequencies and an intensity noise-limited regime at higher Rabi frequencies, which for our phase and intensity noise profiles occurs at  $\Omega_R \approx 2\pi\times7$~MHz. Our numerical simulations suggest that, at Rabi frequencies less than this value, cavity phase noise filtering~\cite{Levine2018} can enhance the long-time Rabi coherence.

\vspace{-3mm}
\section{Rydberg state systematics}\label{sec:rydbergstate}
\vspace{-3mm}
\subsection{State identification and quantum defects}
The Rydberg state $|r\rangle$ we use for this work is the $5$s$61$s~$^3$S$_1~m_\text{J} = 0$ state of $^{88}$Sr. To confirm the quantum numbers, we measure the transition wavelengths of $n = {48,49,50,61}$ for the $^3$S$_1$ series and of $n ={47,48,49}$ for the $^3$D$_1$ series and find nearly perfect agreement with the values predicted by the quantum defects given in Ref.~\cite{Vaillant2012}. 

\begin{figure}[t!]
	\centering
	\includegraphics[width=8.5cm]{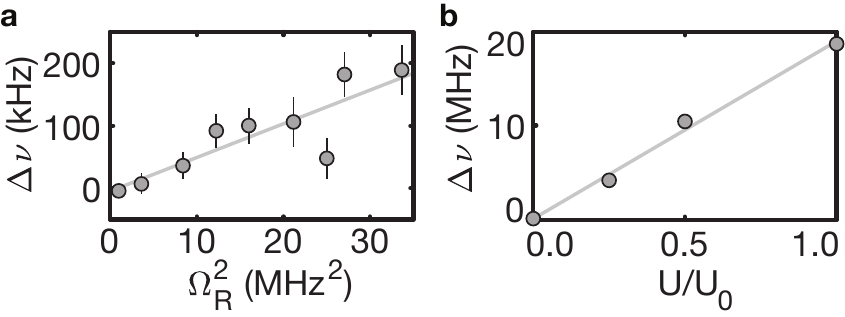}
	\caption{\textbf{Light shifts of $|r\rangle$ from the Rydberg laser and the tweezer light.} \textbf{a}, The differential shift of the $|g\rangle\leftrightarrow|r\rangle$ resonance between $\Omega_{R}^\text{init}=2\pi\times1$ MHz and variable $\Omega_{R}$ versus $\Omega_{R}^2$. This set of data was measured with the two-rail self-comparison technique utilized in Ref.~\cite{Madjarov2019}. The fit line reflects the quadratic scaling $\Delta\nu=\kappa_{|r\rangle}^\text{UV}\Omega_{R}^2$, with $\kappa_{|r\rangle}^\text{UV}=5.1(7)$ kHz/MHz$^2$. \textbf{b}, The differential shift of the $|g\rangle\leftrightarrow|r\rangle$ resonance between the dark case $\text{U}=0$ where the tweezers are extinguished during excitation, and the bright case with variable $|g\rangle$-state depth $\text{U}$ up to $\text{U}_0\approx k_\text{B}\times450$ $\mu$K~$\approx h\times9.4$ MHz. This fit shows a linear dependence with $\Delta\nu=\kappa_{|r\rangle}^\text{T}\text{U}$, where $\kappa_{|r\rangle}^\text{T}=18.8(9)$ MHz/$\text{U}_0$. Error bars indicate a $1\sigma$ standard error of the mean.}\vspace{-0.5cm}
	\label{FigLightShift}
\end{figure}

\vspace{-3mm}
\subsection{Rydberg probe-induced light shift\label{sec:rydbergprobeshift}}
The pulse generation for our Rydberg interrogation is facilitated by switching on and off an acousto-optic modulator (AOM).  However, due to the finite speed of sound in the AOM crystal, the switch-on and switch-off times are limited to tens of nanoseconds. This timescale begins to approach the timescale of our $\pi$-pulses for Rabi frequencies greater than $\approx$10~MHz. This poses a potential problem if there is also a significant intensity-dependent light shift of the resonance frequency due to the Rydberg interrogation beam. For example, a detuning that changes significantly on the timescale of the Rabi frequency could lead to non-trivial dynamics on the Bloch sphere, causing unfaithful execution of Rabi oscillations.  We note that such an effect scales unfavorably with increasing Rabi frequency, as both the relevant timescale becomes shorter and the magnitude of the shift becomes quadratically larger.   

To measure this effect, we operate at Rabi frequencies smaller than 6 MHz to isolate the pure Rydberg probe-induced light shift from any undesired AOM-related transient effects. Using the two-rail self-comparison technique described in Ref.~\cite{Madjarov2019}, we measure the light shift induced by the Rydberg beam and find it to be described by $\Delta\nu=\kappa_{|r\rangle}^\text{UV}\Omega_{R}^2$ with $\kappa_{|r\rangle}^\text{UV}=5.1(7)$~kHz/MHz$^2$, as shown in Fig.~\ref{FigLightShift}a.  

\vspace{-3mm}
\subsection{Tweezer-induced light shift}
We have demonstrated high-fidelity blockaded Rabi oscillations \textit{without} extinguishing the tweezer traps. To gain a partial understanding of this observation, we measure the light shift of $|r\rangle$ in the tweezers with wavelength $\lambda_\text{T}=813.4$ nm and waist of $w_\text{T}\approx800$ nm. We measure the differential shift of the $|g\rangle\leftrightarrow|r\rangle$ resonance between the dark case $\text{U}=0$ where the tweezers are extinguished during excitation, and the bright case with variable $|g\rangle$-state depth $\text{U}$ up to $\text{U}_0\approx450$ $\mu$K~$\approx h\times9.4$ MHz. This fit shows a linear dependence with $\Delta\nu=\kappa_{|r\rangle}^\text{T}\text{U}$, where $\kappa_{|r\rangle}^\text{T}=18.8(9)$ MHz/$\text{U}_0$. We conclude that $\kappa_{|r\rangle}^\text{T}\approx-\kappa_{|g\rangle}^\text{T}$ at this tweezer wavelength and waist. We leave the detailed modeling of the polarizability to future work. 

\vspace{-3mm}
\subsection{Diamagnetic shift from magnetic fields}
We measure a magnetic-field-dependent shift of the Rydberg resonance that is quadratic in the magnitude of the field.  We attribute this shift to the diamagnetic effect~\cite{Weber2017}, which has a Hamiltonian given by ${H_{dm} = \frac{1}{8m_e}|\textbf{d} \times \textbf{B}|^2}$, where $\textbf{d}$ is the dipole operator, $\textbf{B}$ is the magnetic field, and $m_e$ is the electron mass. This Hamiltonian gives rise to a first order shift in the energy that is quadratic in the magnitude of the field such that $\Delta\nu_{dm} = \beta|\textbf{B}|^2$, where $\beta$ is a state dependent quantity that increases with the principal quantum number $n$. For $5$s$61$s~$^3$S$_1~m_\text{J} = 0$, we experimentally measure $\beta \approx 3.4$~kHz/G$^2$.  

We compare this value to a value predicted by performing exact diagonalization of $H_{dm}$ on a limited manifold of Rydberg states in a similar fashion to Ref.~\cite{Weber2017} while using quantum defects from Ref.~\cite{Vaillant2012}.  This numerical procedure produces $\beta_{\text{predicted}} = 2.9$~kHz/G$^2$ for our state, in near agreement with our measured value.   


\bibliographystyle{manubib2}

\end{document}